\documentclass{aastex61}
\submitjournal{ApJ}
\newcommand{\ud}[1]{{{#1}}{}}
\newcommand{\degree}{^{\circ}}
\usepackage{CJK}
\usepackage{xcolor} 
\usepackage{graphicx}
\usepackage{amsmath,bm}
\usepackage{multirow}
\usepackage{array}
\usepackage{lineno}

\begin{document}
\begin{CJK*}{UTF8}{gbsn}

\title{Chromospheric recurrent jets \ud{in a sunspot group and their inter-granular origin}}

\correspondingauthor{Jie Zhao}
\email{zhaojie@pmo.ac.cn}

\author{Jie Zhao}
\affil{Key Laboratory of Dark Matter and Space Astronomy, Purple Mountain Observatory,\\
Chinese Academy of Sciences, Nanjing, Jiangsu, China}

\author{Jiangtao Su}
\affil{Key Laboratory of Solar Activity, National Astronomical Observatories, Chinese Academy of Sciences, Beijing 100012, China}
\affil{School of Astronomy and Space Science, University of Chinese Academy of Sciences, Beijing 101408, China}

\author{Xu Yang}
\affil{Center for Solar-Terrestrial Research, New Jersey Institute of Technology, 323 Martin Luther King Boulevard, Newark, NJ 07102, USA}
\affil{Big Bear Solar Observatory, New Jersey Institute of Technology, Big Bear City, CA 92314, USA}

\author{Hui Li}
\affil{Key Laboratory of Dark Matter and Space Astronomy, Purple Mountain Observatory,\\
Chinese Academy of Sciences, Nanjing, Jiangsu, China}
\affil{School of Astronomy and Space Science, University of Science and Technology of China, Hefei 230026, China}

\author{Brigitte Schmieder}
\affil{LESIA, Observatoire de Paris, Universit\'e  PSL, CNRS, Sorbonne Universit\'e, Universit\'e de Paris, 5 place Jules Janssen, 92195 Meudon, France}
\affil{Centre for Mathematical Plasma Astrophysics, Department of Mathematics, KU Leuven, 3001
Leuven, Belgium}
\affil{LSUPA, School of Physics and Astronomy, University of Glasgow, Scotland}

\author{Kwangsu Ahn}
\affil{Center for Solar-Terrestrial Research, New Jersey Institute of Technology, 323 Martin Luther King Boulevard, Newark, NJ 07102, USA}
\affil{Big Bear Solar Observatory, New Jersey Institute of Technology, Big Bear City, CA 92314, USA}

\author{Wenda Cao}
\affil{Center for Solar-Terrestrial Research, New Jersey Institute of Technology, 323 Martin Luther King Boulevard, Newark, NJ 07102, USA}
\affil{Big Bear Solar Observatory, New Jersey Institute of Technology, Big Bear City, CA 92314, USA}

\begin{abstract}

\ud{We report on high-resolution observations of recurrent fan-like jets by the Goode Solar telescope (GST)  in multi-wavelengths inside a sunspot group.  The dynamics behaviour of the jets is derived from the H$\alpha$ line profiles. Quantitative values for one well-identified event have been obtained showing a maximum projected velocity of 42 $\rm km\ s^{-1}$ and a Doppler shift of the order of 20 $\rm km\ s^{-1}$. The footpoints/roots of the jets have 
a lifted center on the H$\alpha$ line profile
compared to the quiet sun suggesting a long lasting heating at these locations.
The magnetic field  between the small sunspots in the group shows a very high resolution pattern with parasitic polarities along the inter-granular lanes  accompanied by high velocity converging flows  (4 $\rm km\ s^{-1}$) in the photosphere. Magnetic cancellations between the opposite polarities are observed in the vicinity of the footpoints of the jets. Along the inter-granular lanes horizontal magnetic field around 1000 Gauss is generated impulsively.  Overall, all the kinetic features at the different layers through photosphere and chromosphere favor a convection-driven reconnection scenario for the recurrent fan-like jets, and evidence a site of reconnection between the photosphere and chromosphere  corresponding to the inter-granular lanes.}

\end{abstract}


\section{Introduction}\label{sec:intro}

\ud{Recurrent} fan-like jets which appear like a chain of eruptions from one end to another in the high-resolution \ud{chromosphere} observations are \ud{the prevalence of a highly dynamic behaviour.}
Due to the limitation of spatial resolution, they have been named as H$\rm \alpha$ surges, plasma ejections, chromospheric jets \citep[such as in ][]{1973SoPh...28...95R,2001ApJ...555L..65A,2014A&A...567A..96L}. With \ud{the} recent high resolution observations, they have also been named \ud{such as} light wall \citep{2015ApJ...804L..27Y} or peacock jets \citep{2016A&A...590A..57R}. Among such \ud{phenomena,}
the one happens inside the umbra \ud{in} light bridge (LB) has attracted more \ud{attention}. 
The oscillations of fan-like jets (plasma ejections/surges) \ud{inside}
a sunspot LB was first reported in \citet{2001ApJ...555L..65A} with H$\alpha$ observations \ud{by} the Domeless Solar Telescope at Hida Observatory and \ud{171 \AA\ observations from the Transition Region and Coronal Explorer \citep[TRACE,][]{1999SoPh..187..229H}.}
\ud{The main characteristics of these jets were defined by their velocity around 50  $\rm km\ s^{-1}$ and their maximum length of 2Mm.  Emerging magnetic flux was considered to be the origin of the jets with no strong observational evidence.}\\

\ud{Observed in multi-wavelengths}
throughout the solar atmosphere, fan-like jets have been extensively studied in recent years \citep[see the reviews of][]{2018ApJ...854...92T,2021SoPh..296...84D,2021arXiv211109002S}. They are identified in high spatial resolution observations, such as in H$\alpha$ and transition region lines \ud{observed with the} Interface Region Imaging Spectrograph \citep[IRIS,][]{2014SoPh..289.2733D} with different morphologies (velocity, maximum height, period). They appear as dark features \ud{in H$\alpha$ images} with a bright front in the transition region and coronal lines \citep[such as in][]{2015ApJ...804L..27Y}. The jets occur not only above LB but also above magnetic neutral lines \citep{2016A&A...589L...7H}, with trigger of magneto-acoustic waves \citep{2017ApJ...838....2Z}, magnetic reconnection \citep{2017ApJ...848L...9H,2019ApJ...870...90B,2019ApJ...886...64Y} or the combination of both \ud{mechanisms} \citep{2018ApJ...854...92T}. \ud{They may also be associated with vortex flows \citep{2019ApJ...882..175Y}.}\\

The magneto-acoustic wave origins are predominantly identified from the periodic \ud{behaviour of the jet oscillations or of the }displacements of the bright front while the magnetic reconnection origins are confirmed from the high resolution observations of magnetic field. With \ud{the} Hinode spectropolarimeter \citep[SP,][]{2008SoPh..249..233I,2013SoPh..283..579L}, \citet[][hereafter S09]{2009ApJ...696L..66S} found \ud{that a} trapped flux tube beneath the canopy magnetic field is responsible for long-lasting chromospheric plasma ejections. Interaction of magnetic field between LB and the surroundings \ud{is} also suggested to raise the local dynamic \citep{2014A&A...567A..96L,2015ApJ...811..137T}.\\

Simulations of jets, such as X-ray jets \citep{2000ApJ...542.1100S}, surges \citep{1996PASJ...48..353Y}, solar polar jets \citep{2009ApJ...691...61P}, active region recurrent jets \citep{2010A&A...512L...2A}, microflare accompanying jets \citep{2012ApJ...751..152J}, show different physical processes but all \ud{associate} with magnetic field \ud{changes}. \citet{2013PASJ...65...62T} simulated flux emergence associated chromospheric jets and displayed two possible scenarios with different reconnection heights. With a lower reconnection site, such as \ud{one in} the photosphere, a slow-mode shock propagates and \ud{lifts} up the transition region, forming the so-called jets. While with a higher reconnection site, such as \ud{that in} the upper chromosphere, Lorentz force and the whip-like motion of magnetic field accelerate the chromospheric plasma.  
Since the jets originate from LB in some cases, \citet[][hereafter T15]{2015ApJ...811..138T} did a detailed analysis of such structures with MURaM simulation of flux emergence in an active region. 
Their results show that the convective upflow transport horizontal field to the solar surface, configuring light bright structure, and the \ud{jet-associated} magnetic reconnection happens due to the magnetic shear between the horizontal field of the light bridge and the ambient vertical field in the sunspot. The morphology, magnetic field as well as convective velocity near the solar surface are compared between the Hinode observations and MHD simulation around the LB, which shows a good correspondence.\\

Although previous observations have shown the intimate relation between the fan-like jets and the photospheric magnetic field, the spatial resolution is limited to subsecond, such as $\rm 0$\arcsec$.3\ pixel^{-1}$ of Hinode/SP, and a detailed comparison between different layers is rarely achieved.
With the aid of extremely high resolution observations from Goode Solar Telescope \citep[GST,][]{2010AN....331..636C}, we associate the chromospheric plasma eruptions with the photospheric granule movements as well as the inter-granular magnetic field. Our paper is organized as follows: In section \ref{sec:obs}, we list the observations that we investigated in this work and the inversion method that adopted for obtaining the magnetic field.
The results are shown in Section \ref{sec:results} and we discuss our conclusions in Section \ref{sec:conclusion}.\\

\section{observations and data reduction}\label{sec:obs}
Intermittently recurrent fan-like jets which happened in between the well-developed east sunspot of NOAA Active Region 12585 ($\rm N07\degree,W25\degree$) have been observed in multi-wavelengths with GST at the Big Bear Solar Observatory (BBSO) on 2016 September 7. 

The H$\alpha$ 
\ud{images are} observed by the Visible Imaging Spectrometer \citep[VIS,][]{2010AN....331..636C}, with a scanning wavelength step 0.02 nm from -0.12 to 0.12 nm of \ud{the} line center. The \ud{pixel size} is $\rm 0$\arcsec$.029$, with a effective temporal resolution of 33 seconds for every single wavelength.

The TiO \ud{images} at 705.7nm \ud{are} observed by the Broadband Filter Imager \citep[BFI,][]{2010AN....331..636C}, with a bandpass of 1 nm, showing the photospheric counterpart of the jets. The temporal resolution is 30 seconds and the \ud{pixel size} is $\rm 0$\arcsec$.034$.

Full-Stokes Near Infra-Red Imaging Spectropolarimeter \citep[NIRIS,][]{2012ASPC..463..291C,2016SPD....47.0207A,2019ASPC..526..317A} observes the photospheric line of Fe I 1564.85 nm for magnetic stokes profiles, with a \ud{pixel size} amount to $\rm 0$\arcsec$.083$. There are 40 spectral sampling positions from -0.316 to 0.31 $\rm nm$ with respect to the line center, and the cadence is around 73s for vector magnetic field.

For the inversion of magnetic field and other parameters from the stokes profiles, a Milne-Eddington atmosphere is assumed, such as the one in \citet{2017NatAs...1E..85W}, and a Minimum-Energy Approach \citep{2009ASPC..415..365L,2009SoPh..260...83L} is adopted to resolve the 180 degree ambiguity in the azimuth angle of the vector magnetic field.

For the alignment of images in different wavelengths, the \ud{TiO} images are rotated and shifted with respect to the \ud{continuum image of Helioseismic and Magnetic Imager \citep[HMI,][]{2012SoPh..275..207S}}, then the images of stokes profiles and H$\alpha$ far wings are co-aligned with the TiO images, and finally the H$\alpha$ near line center images are co-aligned with the H$\alpha$ far wings images. 
We show the result after co-alignment (all north-up) at one time step in Figure 1 when the \ud{jet is} apparent in the field-of-view (FOV) which covers a region of 60$\rm \arcsec\times 60\arcsec$. The photospheric \ud{image} in TiO band and HMI continuum image are displayed in the upper panels and the chromospheric responses in the red wings of H$\alpha$ are shown in the lower panels. The fan-like jets are apparent in between the sunspots and have been annotated with blue curves along the ejection trajectory.\\

\section{results}\label{sec:results}
\subsection{chromospheric jets}\label{subsec:jets}
\begin{figure}[htbp]
\centering
\includegraphics[width=6cm, trim=10cm 0cm 10cm 0]{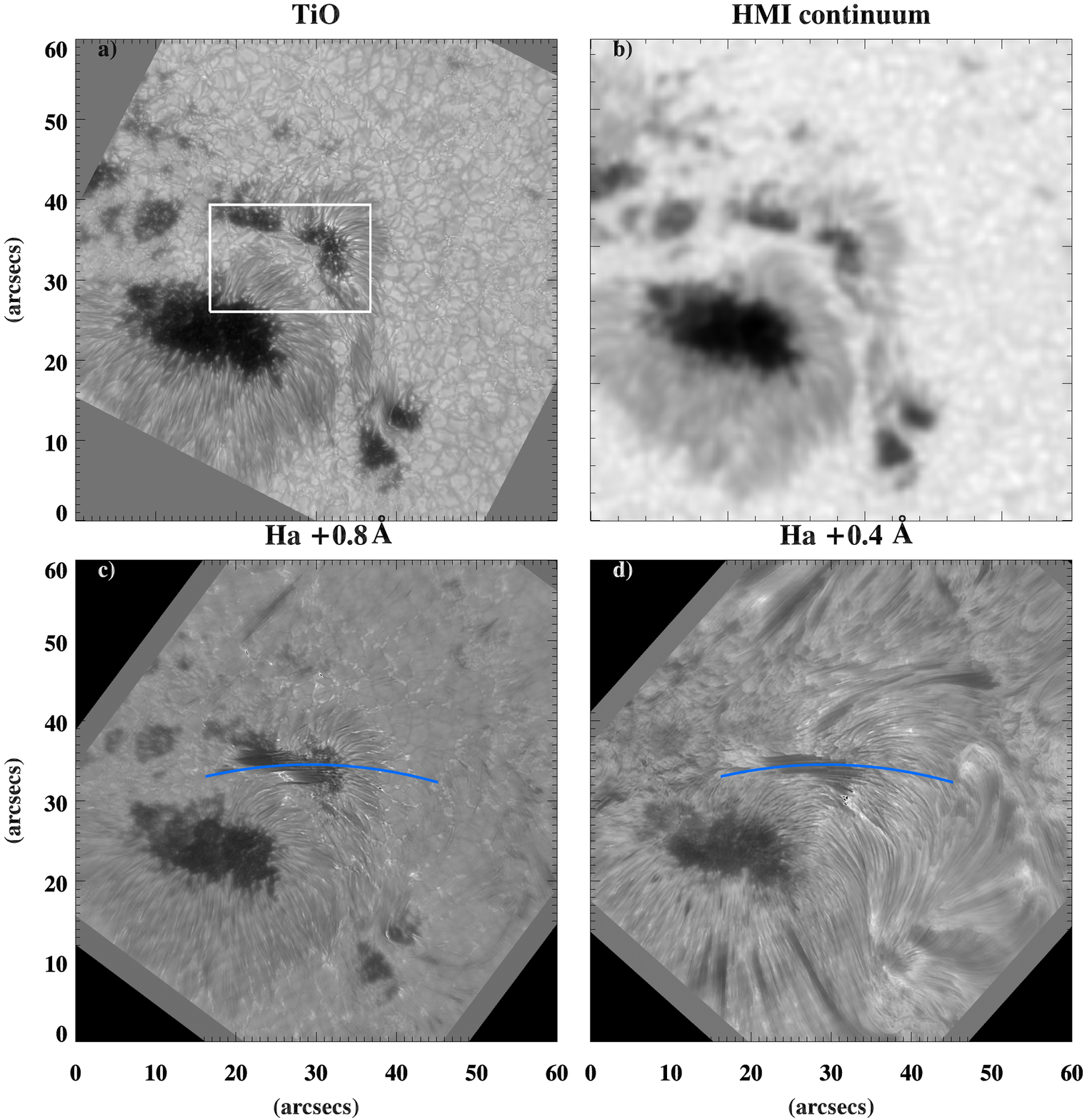}
\caption{Overview of the coaligned observations \ud{of} AR12585: a) TiO white-light (705.7nm) image from GST/BFI; b) SDO/HMI continuum (617.3nm) image; Chromospheric images observed in c) H$\alpha$ + 0.8$\rm \AA$ and d) H$\alpha$ + 0.4$\rm \AA$ by GST/VIS. All the images from GST are rotated and coaligned with the HMI continuum \ud{image}. The white rectangle box in TiO image shows a FOV displayed in Figures \ref{fig:ha_tio} and \ref{fig:niris_tio}. The recurrent fan-like jets \ud{are} identified from the chromospheric images in the lower panels and their trajectories are annotated with blue curves. }\label{fig:obs_overview}
\end{figure}

\begin{figure}[htbp]
\centering
\includegraphics[width=14cm,trim=3cm 2.5cm 3cm 2.5cm]{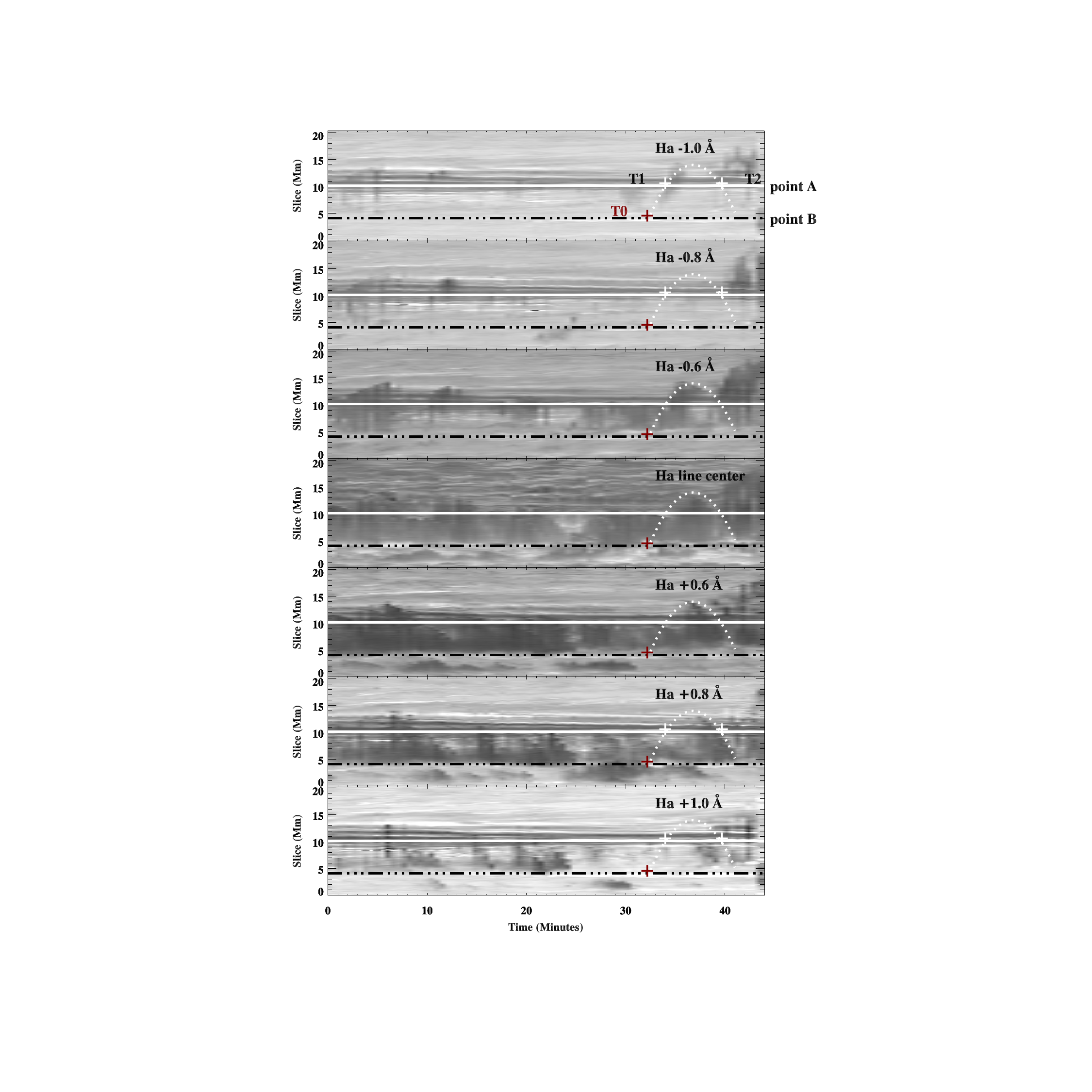}
\caption{Time-distance plots that sample the recurrent fan-like jets along the blue curve in Figure \ref{fig:obs_overview} \ud{are displayed with a time range from 17:00 UT to 17:44 UT.} Evolution in the wavelengths \ud{from H$\rm \alpha - 1.0\ \AA$ to H$\rm \alpha + 1.0\ \AA$ is shown from top to bottom. One jet event has been selected and outlined with white dotted parabolic lines. Point A and B, which are annotated with white line and black dash-dotted line, are selected to represent a point in the jet and the root, respectively. The white plus signs in the top and bottom two panels show the times (T1 and T2) when upflow and downflow, which are mainly observed in H$\rm \alpha \pm 0.8\ \AA$ and $\rm \pm\ 1.0\ \AA$, is passing through point A, respectively. The red plus sign in all panels shows the time (T0) when the root of the selected jet event is investigated later. The intensity evolution at point A and B of H$\rm \alpha$ from the blue wing to the red wing with a wavelength step of $\rm 0.2\ \AA$ are stacked and displayed in Figure \ref{fig:ha_spectro}. } }\label{fig:time_dis}
\end{figure}

To investigate the \ud{temporal} evolution of the fan-like jets, \ud{we consider} the intensity at H$\alpha$ line \ud{center and} wings along the blue curve in Figure \ref{fig:obs_overview} stacked over time and the time-distance maps shown in Figure \ref{fig:time_dis}. \ud{A time range of 17:00 UT -- 17:44 UT is selected} considering the overlap of the multi-wavelengths observations as well as the data quality. \\

Fan-like jets appear as dark features in the time-distance map while the background of the chromosphere is bright. The intensity fluctuation of the jets indicates upward movements in \ud{the blue wings (H $\rm \alpha - 0.6,0.8,1.0 \AA$)} and the downward movements can also be identified in the red wings \ud{(H $\rm \alpha + 0.6,0.8,1.0 \AA$).} Nevertheless, the dark plasma jets are less intense in the far wings of H $\rm \alpha\ \pm$ 0.8, \ud{1.0$\AA$}, especially in the blue wing, which manifests asymmetric distribution of the upward and downward movement of the jets at high speed. \ud{One of the recurrent jets which can be clearly identified from the time-distance map of H $\rm \alpha \pm 0.8$ and $\rm 1.0 \AA$ has been selected and its trajectory has been fitted with a parabolic profile \ud{following the method used in \citet{2016A&A...590A..57R}}. The averaged projected velocity is then  estimated to be 42 $\rm km\ s^{-1}$. For the rest ejections,} most of them can not be well-established due to the blend of the upflow with downflow. \\

\begin{figure}[htbp]
\centering
\includegraphics[width = 14cm,trim= 3cm 3.cm 3cm 3.cm, clip]{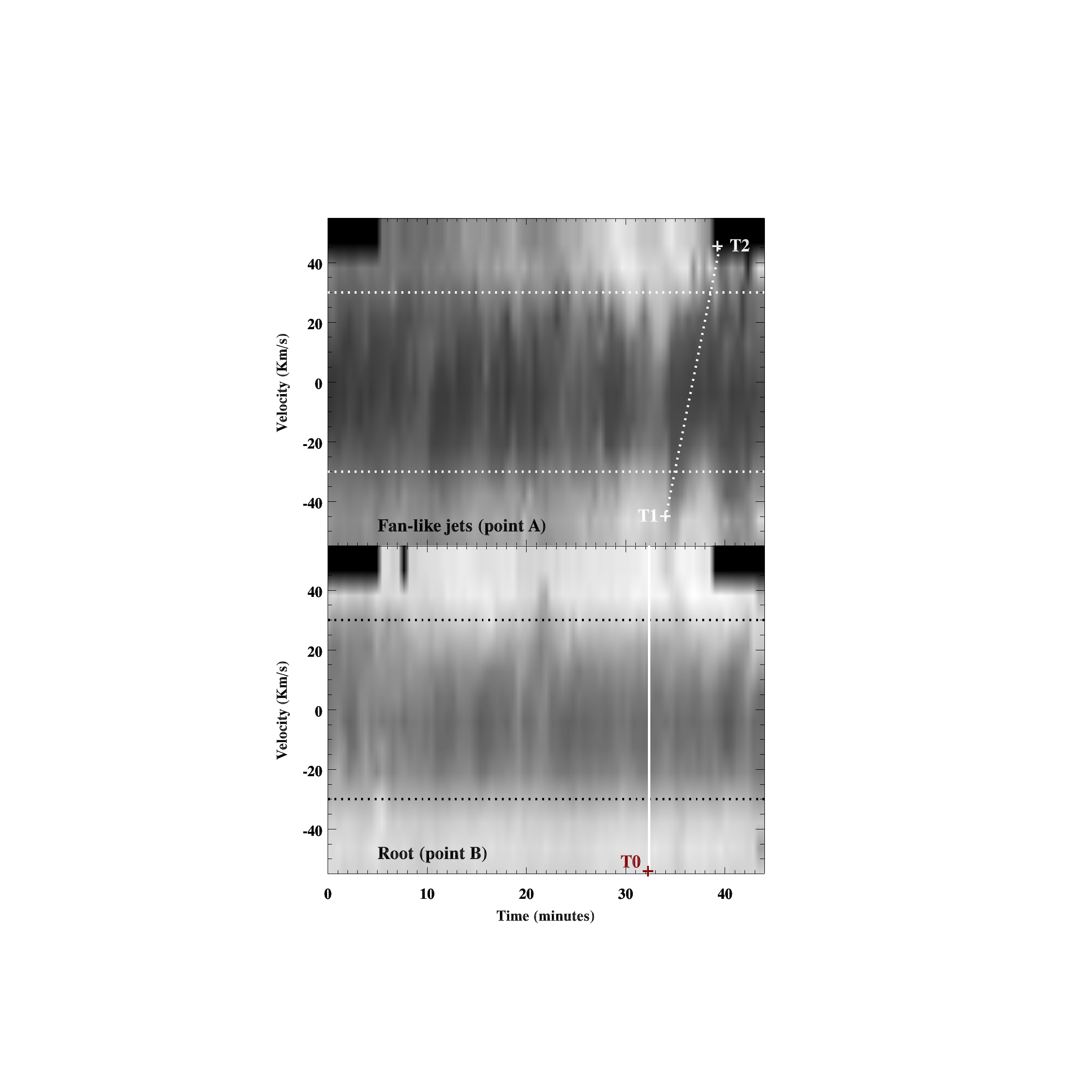}
\caption{Temporal evolution of stacked H$\alpha$ intensity from blue wing to the red wing for \ud{point A on the path of the fan-like jets (top panel) and point B of the root (bottom panel).} Different wavelengths along the vertical axis have been transformed into Doppler velocity, and the ones of $\rm \pm\ 30\ km\ s^{-1}$ are annotated with dotted horizontal lines \ud{for reference. The oblique dotted line in the top panel demonstrates the same jet event and the white plus signs at the two ends mark the same times (T1 and T2) as demonstrated in Figure \ref{fig:time_dis}. Both the vertical white line and the red plus sign in the bottom panel show the time (T0) when the intensity profile of the root is displayed in Figure \ref{fig:ha_spectro_heating}.} Dark blocks at the corners exist due to the missing observations at H$\alpha$ far wing in red.}\label{fig:ha_spectro}
\end{figure}

Two points \ud{of A and B} on the slice are selected and marked with horizontal white solid line and black dash-dotted line in Figure \ref{fig:time_dis} to represent the jets and the chromospheric roots respectively. \ud{The two crossing points of the white horizontal line and the parabolic profile are marked with white plus signs to represent the times when upflow (at time T1) and downflow (at time T2) are passing through point A consecutively during the selected jet event. The red plus sign at the first crossing point of the dash-dotted line and the parabolic profile is annotated to show the time (T0) when the root is intimately associated with the selected jet event and is investigated later. Examples of} temporal evolution of the H$\alpha$ intensity from blue wing to the red wing along the aforementioned two horizontal lines \ud{are stacked} and displayed in Figure \ref{fig:ha_spectro}.
Top panel shows the result for the fan-like jets and bottom panel for the root. The wavelength in the vertical axis is displayed in the format of Doppler velocity, which is calculated with $v = ({\delta\lambda}/{\lambda_{0})\cdot c}$, where $\lambda_{0}$ is the line center wavelength of H$\alpha$ and $\delta\lambda$ is the offset to the line center. \ud{In general, intermittent event can be found in both panels while no correspondence can be related with most of the fan-like jets and the root. This is probably due to the mixture of the jets between two consecutive eruptions, i.e. the coordinated upflow and downflow which also complicate the Doppler velocity distribution. The dark feature of the fan-like jets and the root mainly locates in between $\rm \pm 30\ km\ s^{-1}$ (white dotted lines), showing recurrent blue shift to red shift pattern. The identified jet event in Figure \ref{fig:time_dis} is also marked with a dotted line in the top panel, with T1 and T2 show the times when the upflow and downflow passing through the point A during the jet event. The Doppler velocity shows a response in all the blue wing of H$\alpha$ at the same time while the response in the red wing appears consecutively from the near line center to the far wings. The result of the rapid blue shift followed by a slowly increasing red shift is the manifestation of the so-called magnetoacoustic shock waves \citep[e.g.,][]{2016ApJ...817..117S}. The time (T0) when the root of the selected jet is studied is also marked with a red plus sign as in Figure \ref{fig:time_dis} and a vertical line in the bottom panel.} 
\\

\begin{figure}[htbp]
\centering
\includegraphics[width = 10cm,trim=4cm 5.5cm 2cm 8cm]{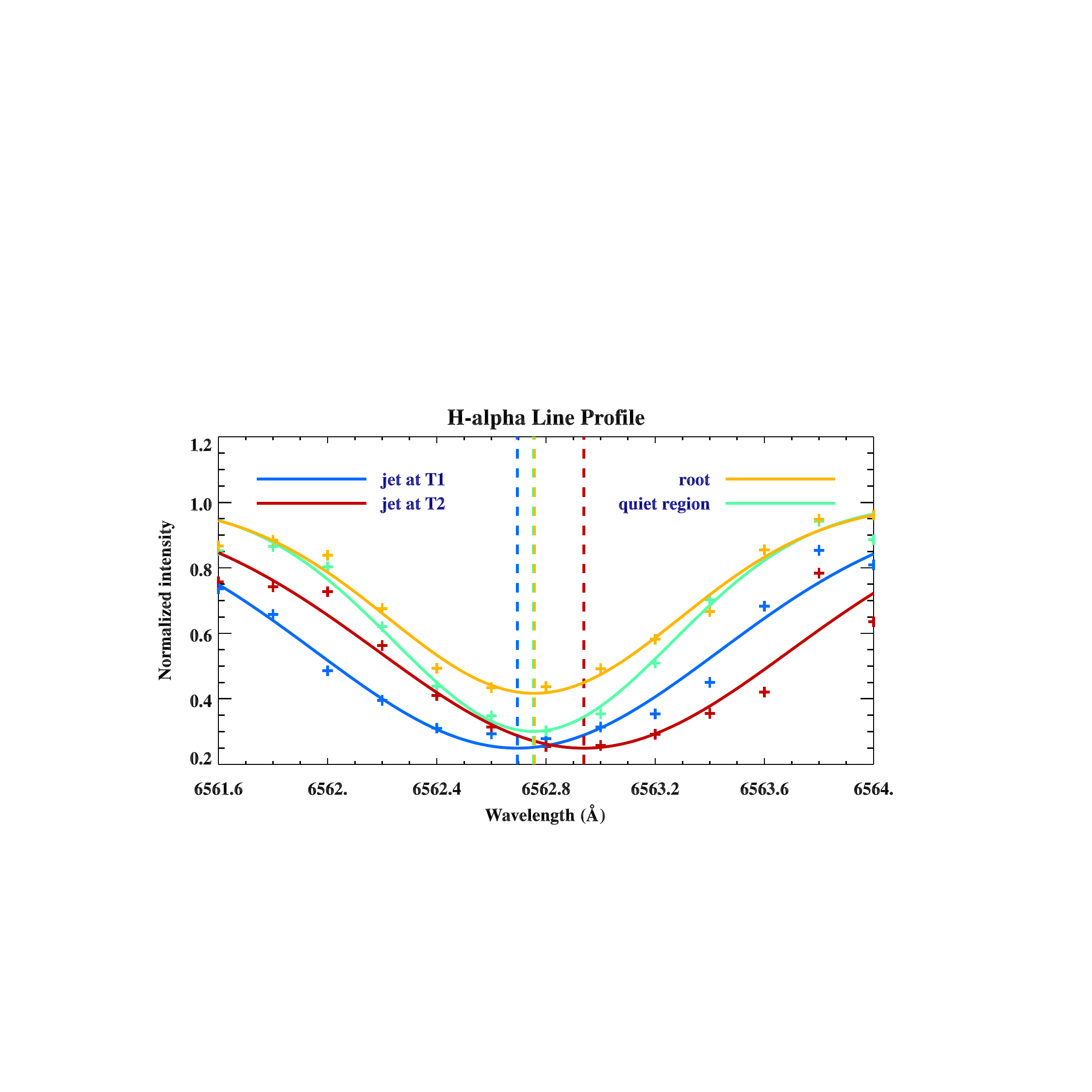}
\caption{\ud{Normalized intensity profiles versus wavelengths for the jet (at T1 and T2), the root (at T0) and the quiet region (at T0). The plus signs show the observed data sets and the line curves show the fitting results. The vertical dotted lines give the line center from the fitted curves.}}\label{fig:ha_spectro_heating}
\end{figure}

The normalized H$\rm \alpha$ line profiles \ud{at time T0 for the chromospheric root and quiet region, and at times T1 and T2 for the jet are shown in Figure \ref{fig:ha_spectro_heating}. The plus signs with different colors show the observed data sets for different features while the curves show the fitting results. The line centers from the fitted results are labeled with vertical lines. The line profile at the chromospheric quiet region (green line) shows a blue shift which corresponds to a velocity around 2 $\rm km\ s^{-1}$ if the line center of H$\alpha$ is assumed to be 6562.8 $\rm \AA$. The H$\rm \alpha$ line center of the jet shows blue shift at T1 ($\rm\sim 3\ km\ s^{-1}$) and red shift at T2 ($\rm\sim 8\ km\ s^{-1}$), while the one of the root (yellow profile) does not show apparent Doppler shift, both relative to the quiet region. A cloud model analysis following the formula of \citet{1988A&A...201..327S} may lead to higher Doppler shifts for the jet at times T1 and T2 of the order of 15 to 20  $\rm km\ s^{-1}$, giving an estimation of the inclination of the jet versus the vertical $\sim 60 \degree$.
Comparing with the quiet region, less absorption is found around the line center for the root while more absorption is found for the jet, which means \ud{that} the root might be heated while the jet is dominated with cold plasma without apparent heating. Although less absorption of H$\alpha$ does not always mean local heating, it is indeed a manifestation of the reconnection that happens in between the photosphere and chromosphere in this study, which is approved later and considered to be responsible for the recurrent jets.}\\

To investigate the photospheric origins of the fan-like jets, a subregion (white rectangle box in Figure \ref{fig:obs_overview}) \ud{was} selected. \ud{The images of H$\alpha$ line center (top row) and line wing at H$\rm \alpha - 0.8\ \AA$ (bottom four rows) at three time steps around the selected jet event shown in Figure \ref{fig:time_dis} and \ref{fig:ha_spectro} are displayed in Figure \ref{fig:ha_tio}. Different features of surface velocity, horizontal magnetic field and longitude magnetic field are overlaid on the H$\rm \alpha - 0.8\ \AA$ images in the bottom three rows, respectively. In the top two rows, the recurrent jets are found to be originated from the same places and two regions of FOV1 and FOV2 as labeled in the middle three rows are selected for investigation later. \ud{The red arrows in each panel show the locations of opposite polarities which corresponds to the inter-granular lanes
in Figure \ref{fig:niris_tio}.}} The \ud{white and black} arrows in the \ud{third} row show the surface flows \ud{in between the magnetic polarities} that \ud{are} obtained from TiO images with Local Correlation Tracking (LCT) method, and their colors represent surface flow with positive and negative $\rm B_{l}$ respectively. \ud{Convective flows larger than 2 $\rm km\ s^{-1}$ (with maximum around 4 $\rm km\ s^{-1}$) exist in FOV2 while flows in FOV1 are relative small. 
An enlargement of the above velocity can also be found in the zoom-in plot in Figure \ref{fig:niris_tio}, in which the converging motions to the inter-granular lanes are identified at some places in the vicinity of the roots. Such flows may help to squash the magnetic field of opposite polarities and trigger the magnetic reconnection.} The horizontal magnetic field $\rm B_{h}$ which is \ud{smaller than 1200 G with $\rm B_{l}$ less than 1000 G has been overlaid on the fourth row, and the longitude magnetic field $\rm B_{l}$ which has an absolute value in the range of 100 -- 800 G with $\rm B_{h}$ less than 1000 G has been overlaid on the fifth row.} The former one is represented with arrows (the pink and green colors show $\rm B_{h}$ with positive and negative $\rm B_{l}$ respectively) and the latter one is shown with circles (white and black represent positive and negative $\rm B_{l}$). 
The magnetic field around the footpoints shows \ud{opposite polarities of $\rm  B_{l}$}, which highly indicates a magnetic field origin \ud{for the jets.} A detailed description of the obtained magnetic field is shown in  \ref{subsec:mag}.\\

\subsection{magnetic fields and \ud{Doppler velocity} from inversion}\label{subsec:mag}
\begin{figure}[htbp]
\centering
\includegraphics[width = 17cm,trim=0cm 0cm 0cm 0cm]{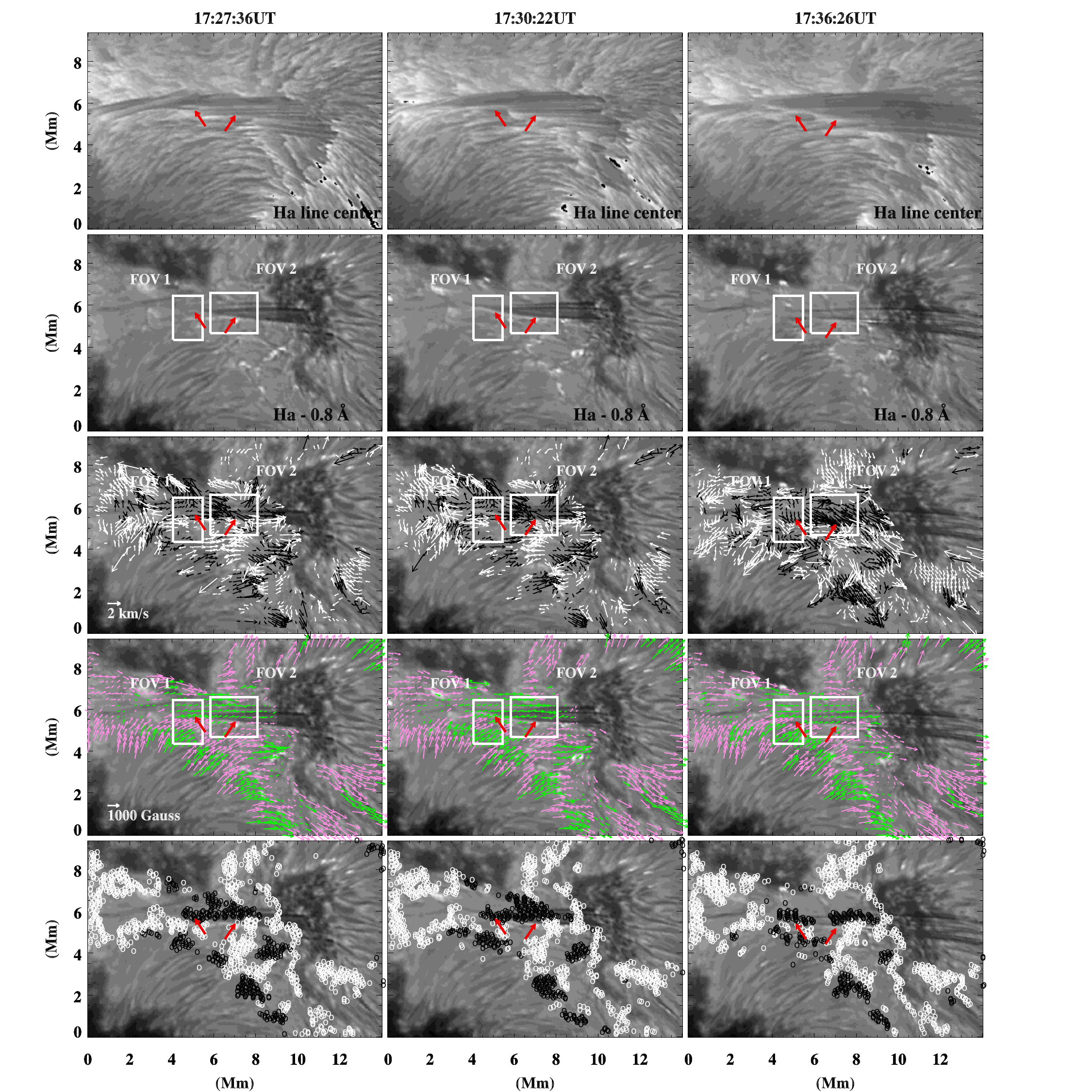}\\
\caption{\ud{The images of H$\rm\alpha$ line center (first row) and H$\rm \alpha - 0.8\ \AA$ (bottom four rows) at three different times are displayed from left to right, different features are overlaid on the H$\rm \alpha - 0.8\ \AA$ images in the bottom three rows. Third row:} The surface (horizontal) velocity at photosphere is overlaid with arrows, and \ud{white and black} colors correspond to velocity with positive and negative $\rm B_{l}$, respectively. \ud{Fourth} row: The horizontal magnetic field is overlaid with arrows, while pink and green colors corresponds to $\rm B_{h}$ with positive and negative $\rm B_{l}$, respectively. \ud{Fifth} row: The longitude magnetic field in the range of 100 -- 800 G with $\rm B_{h}$ less than 1000 G is overlaid with circles , and white and black colors show positive and negative $\rm B_{l}$. \ud{The red arrows indicate the locations of opposite polarities which \ud{correspond} to the inter-granular lanes in Figure \ref{fig:niris_tio}, the white boxes \ud{marked with} FOV1 and FOV2 show the places where the magnetic field is investigated (see Section \ref{subsec:mag_evo}).}
}\label{fig:ha_tio}
\end{figure}

\begin{figure}[htbp]
\centering
\includegraphics[width = 17cm,trim=0cm 0cm 0cm 0cm]{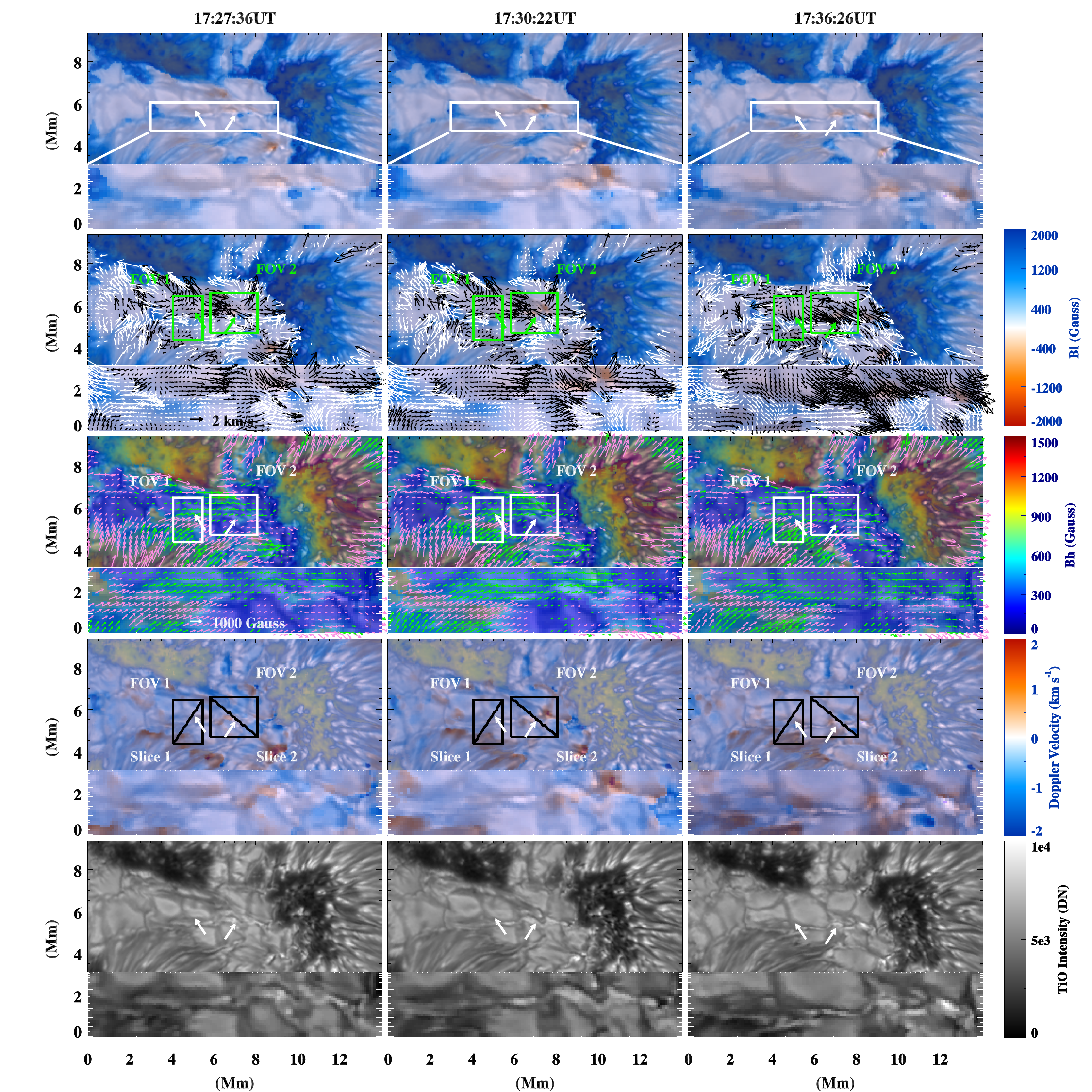}\\
\caption{Composite images of TiO with LOS magnetic field ($\rm B_{l}$, first two rows), horizontal magnetic field ($\rm B_{h}$, third row) \ud{and Doppler velocity ($\rm V_{dop}$, fourth row) are displayed in the top four rows while the TiO images are displayed in the bottom for comparison.} The evolution at the same time steps as in Figure \ref{fig:ha_tio} \ud{is} shown from left to right. \ud{The FOV in the white rectangle box as shown in the first row is enlarged and the zoom-in is displayed for each panel at its bottom. }The colorbars on the right hand which are for the \ud{three} parameters, i.e., $\rm B_{l}$, $\rm B_{h}$, $\rm V_{dop}$ are displayed in the case of zero transparency. Surface velocity and horizontal magnetic field are overlaid as arrows on the images of $\rm B_{l}$ and $\rm B_{h}$ in the \ud{second and third row, respectively.} The colors of arrows have a same meaning as shown in Figure \ref{fig:ha_tio}. Slice 1 and Slice 2 at the diagonal of FOV1 and FOV2 respectively show the places where time-distance map in Figure \ref{fig:time_slice1} and \ref{fig:time_slice2} are done, and the rectangle boxes of FOV 1 and FOV 2 show the regions where the parameters in Figure \ref{fig:params_evo} are calculated. \ud{The white arrows (green in second row) mark the locations of inter-granular lanes as shown in Figure \ref{fig:ha_tio}.}}\label{fig:niris_tio}
\end{figure}

The photospheric vector magnetic field as well as \ud{the Doppler velocity are} obtained through inversion of \ion{Fe}{1} Stokes profiles. Their composite images with TiO are shown in \ud{the top four rows in} Figure \ref{fig:niris_tio} \ud{while the TiO images are displayed in the bottom for comparison. The images have the same field-of-view as in Figure \ref{fig:ha_tio} and temporal} evolution corresponding to the displayed time steps in Figure \ref{fig:ha_tio} is shown from left to right. \ud{An enlargement of the FOV in the white rectangle box labeled at the first row is displayed for each panel at its bottom part.}\\ 

In general, at the region in between the sunspots with same positive polarity, the magnetic field \ud{is complex. At the outer side of the penumbra of the main pore at the left corner, the horizontal field $\rm  B_{h}$ is around 1000 G and the longitudinal magnetic field $\rm  B_{l}$ shows a mixed pattern of positive and negative. The Doppler velocity shows red shift which is the manifestation of the Evershed downflow. At the rest place in between the sunspots, the absolute value of the longitudinal magnetic field $\rm B_{l}$ is less than 100 G and the horizontal field $\rm  B_{h}$ is less than 300 G, with absolute value of Doppler velocity less than $\rm 0.5\ km\ s^{-1}$ at most places. However, relative strong magnetic field of $\rm  B_{l}$ and $\rm  B_{h}$, surface convective flow as fast as 4 $\rm km\ s^{-1}$ and Doppler velocity as large as $\pm$2 $\rm km\ s^{-1}$ appear \ud{inside the enlarged region} at the boundaries of granules, 
\ud{where the convective flow shows converging motions to the inter-granular lanes.}}\\

\ud{The jet threads are developed as a fan with footpoints following  a  brighter line in the south of the well visible threads in H$\rm \alpha$ line center (Figure \ref{fig:ha_tio}, top panels). We concentrate our study to some of them. They correspond to the inversion line of the magnetic field (also the inter-granular lanes) indicated by the red arrows in Figure \ref{fig:ha_tio}.}
From the distribution of the above four parameters \ud{around the inter-granular lanes,} we notice \ud{that} there are two regions (FOV1 and FOV2 as annotated in Figure \ref{fig:niris_tio}) with distinct properties. In the region of FOV1, the magnetic field is dominated by $\rm B_{h}$ and the Doppler velocity at the inter-granular lanes \ud{is} blue shift dominated, while in the region of FOV2, the magnetic field is dominated by $\rm B_{l}$ and the Doppler velocity at the inter-granular lanes \ud{is} red shift dominated. 
\ud{The inter-granular lanes usually have Doppler red-shift (downflows) and longitude-dominated magnetic field at the quiet region while Doppler velocity and the magnetic field may have different features \ud{if there is flux emergence or energy release.} 
Flux emergence on granular scale observed by the New Vacuum Solar Telescope with high resolution \citep{2022ApJ...925...46S} shows that the flux emerges as dark patch like inter-granular lane and the associated surge has footpoints closely rooted in these inter-granular lanes, exhibiting horizontal magnetic field and convergence flows. Elongated horizontal magnetic field has also been found during the emergence of the top of small loops \citep{2018ApJ...856..127G}.}
For understanding their different roles for driving the recurrent fan-like jets, we do a time-distance investigation along \ud{Slices} 1 and 2 as labeled in bottom panels of Figure \ref{fig:niris_tio}, both of the slices pass through the inter-granular lanes \ud{in the vicinity of the fan-like jets.} \\

\subsection{\ud{magnetic cancellation}}
\begin{figure}[htbp]
\centering
\includegraphics[width = 17cm,trim=0cm 7cm 0cm 0cm]{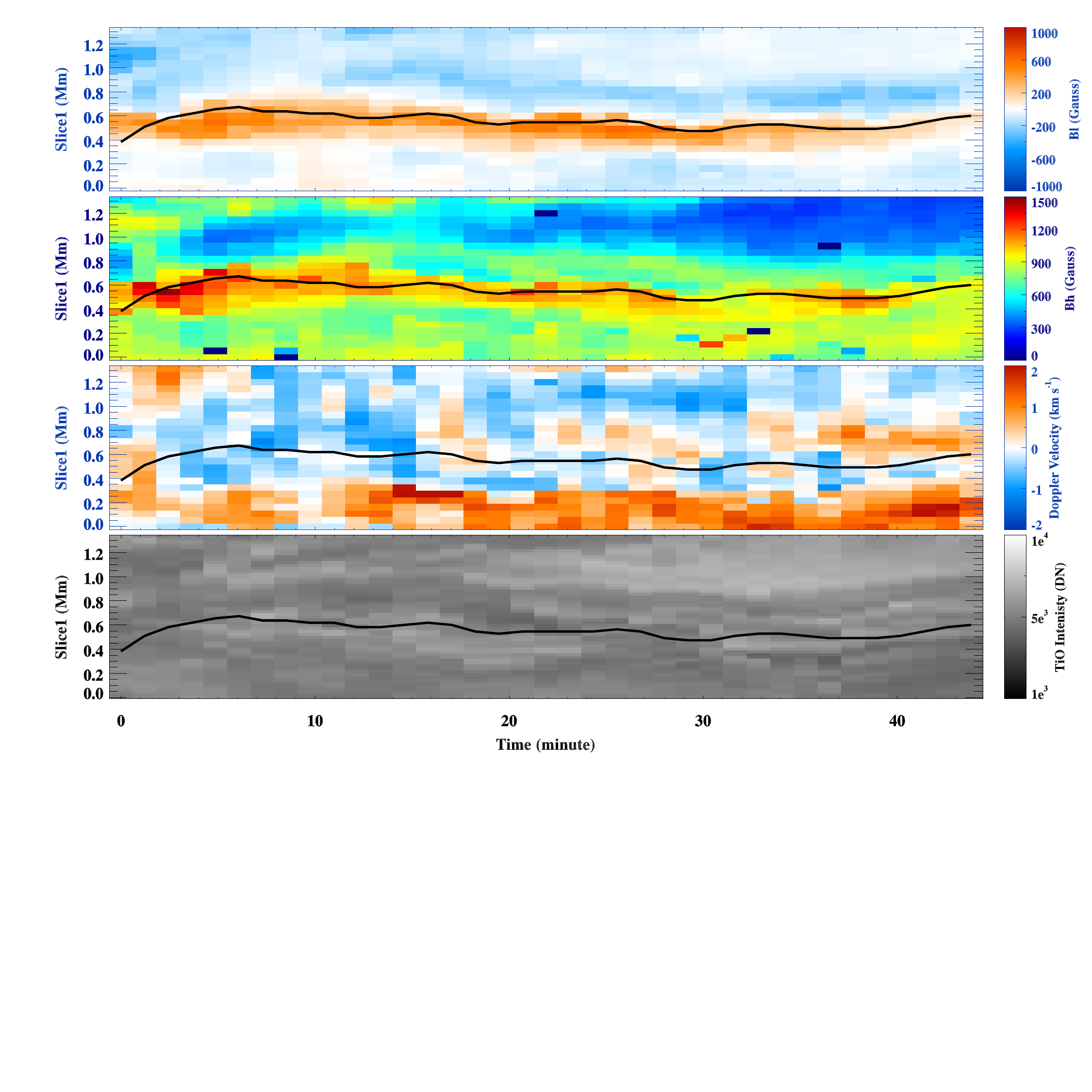}\\

\caption{Time distance images of different parameters at Slice 1-- from top to bottom are: longitude magnetic field, horizontal magnetic field, doppler velocity and TiO intensity. The black curve is obtained by select the maximum horizontal field at each time step and is plotted on the other three panels for reference.
}\label{fig:time_slice1}
\end{figure}

\begin{figure}[htbp]
\centering
\includegraphics[width = 17cm,trim=0cm 7cm 0cm 0cm]{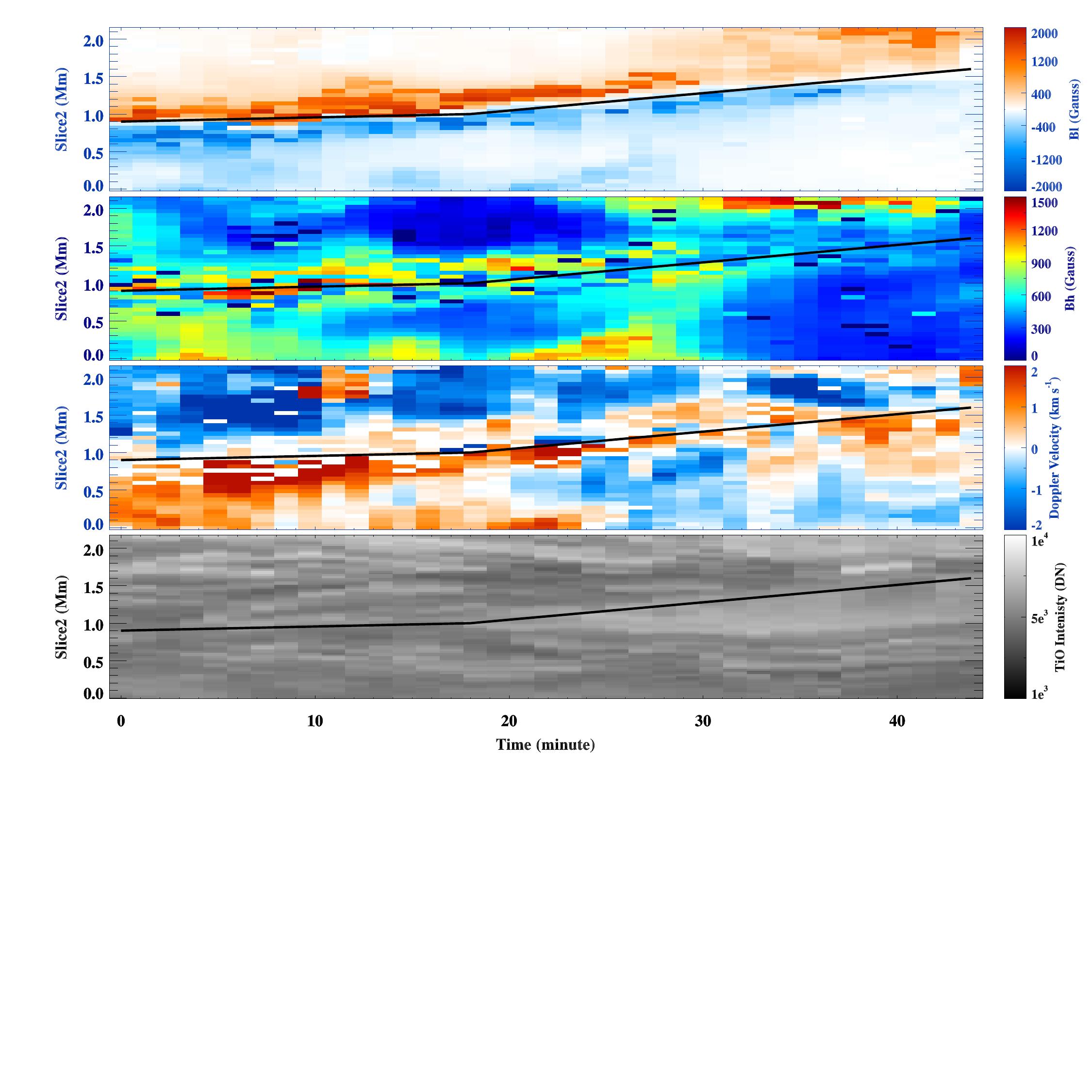}\\

\caption{Same as in Figure \ref{fig:time_slice1} but for Slice 2. The black curve is plotted to represent the place in between the opposite longitude magnetic field and is plotted on the other three panels for reference. 
}\label{fig:time_slice2}
\end{figure}

Time-distance images of Slices 1 and 2 are shown in Figure \ref{fig:time_slice1} and \ref{fig:time_slice2}, respectively. The longitude magnetic field, horizontal magnetic field, Doppler velocity and TiO intensity are displayed from top to bottom. 

In Figure \ref{fig:time_slice1}, the black line curve is selected, according to the maximum $\rm B_{h}$ at each time step, to represent the inter-granular lanes.
\ud{The horizontal field is always high at the inter-granular lanes where opposite polarities are identified and the Doppler velocity shows a blue shift.}
The TiO intensity near the black line curve is higher than its surroundings at most time steps, indicating the existence of heating at photosphere. 

In Figure \ref{fig:time_slice2}, the black line curve is selected manually to represent the inter-granular lanes.  From the first two rows, \ud{an impulsive cancellation process is apparent}, i.e., the opposite polarities of $\rm B_{l}$ decreases while the $\rm B_{h}$ increases intermittently. More evidence \ud{is} found from the bottom two rows, where the enhancement of TiO images \ud{is} identified near the black line and the Doppler velocity is red-shift dominated there, giving hint of magnetic cancellation downflow. \ud{The feature of downflow at the photosphere indicates the reconnection happens above, which is \ud{common} for the reconnection between longitude-component dominated magnetic field with opposite polarities. The impulsive cancellation is also consistent with the relative large convective flow as mentioned above. The TiO intensity at the inter-granular lane is not always bright, as the magnetic cancellation happens intermittently and the energy release might also be disturbed by other activities (such as flux emergence or granule convection adjacent).}\\

\subsection{Temporal evolution of magnetic field}\label{subsec:mag_evo}
\begin{figure}[htbp]
\centering
\includegraphics[width = 14cm,trim=2cm 0cm 2cm 0cm]{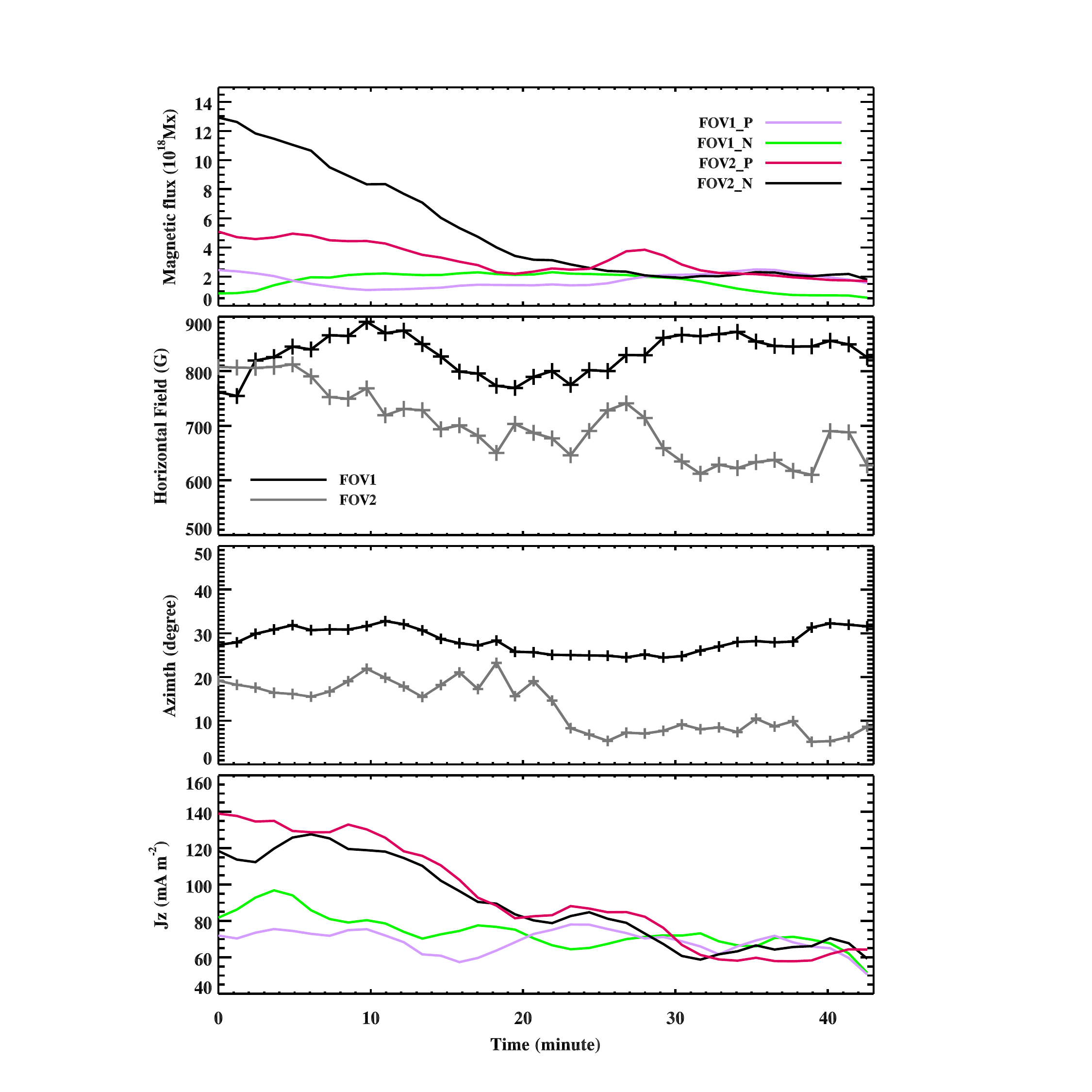}

\caption{Evolution of magnetic features at two regions as shown in Figure \ref{fig:niris_tio}. From top to bottom are magnetic flux (positive and absolute value of negative), mean horizontal field, mean azimuth and mean vertical current density (positive and absolute value of negative), respectively. The legends of different colors in the panels of magnetic flux and horizontal magnetic field are also applicable to the vertical current density and the azimuth, respectively. 
}\label{fig:params_evo}
\end{figure}

To show an overall magnetic properties at the above mentioned two regions of FOV1 and FOV2, four parameters are calculated. Evolution of the magnetic flux, the mean horizontal field, the mean azimuth and the mean vertical current density are plotted from top to bottom panels in Figure \ref{fig:params_evo}. The absolute values of magnetic flux of positive and negative $\rm B_{l}$ in the two regions are shown in the first panel. Both magnetic fluxes of positive and negative polarities in FOV2 show a decay tendency (from -1.3$\rm \times10^{19}$ to -2$\rm \times10^{18}$ Maxwell for negative flux and from 5$\rm \times10^{18}$ to 2$\times10^{18}$ Maxwell for positive flux) while they do not change much in FOV1. The mean value of $\rm B_{h}$ and azimuth of vector magnetic field in FOV1 show variations \ud{with} time but have less \ud{intermittent} change compared to the ones in FOV2. 
The horizontal magnetic field in FOV2 after reconnection is more parallel to the inter-granular lanes which are almost along the X-axis as shown in Figure \ref{fig:niris_tio}. 
Both vertical current density of positive and negative have apparent decrease in FOV2 (from 140 to 60 $\rm mA\  m^{-2}$ for positive current density and from -120 to -60 $\rm mA\  m^{-2}$ for negative current density) while slight decay \ud{is} found in FOV1. 

\ud{In general, the magnetic cancellation is apparent in FOV2, while no clear evidence \ud{is} found in FOV1. As the field with opposite polarities have been identified at the inter-granular lanes in Figure \ref{fig:time_slice1}, the magnetic cancellation is preferred to happen. However, as the magnetic flux in FOV1 shows slightly increase for the positive polarity around 17:30UT, magnetic flux emergence is also suspected to happen at the inter-granular lanes where the Doppler velocity is blue shift as demonstrated in Figure \ref{fig:time_slice1}. Such magnetic emergence might be co-spatial with the magnetic cancellation hence weaken the performance of the latter one in FOV1.}\\

\section{Discussion and Conclusion}\label{sec:conclusion}
\ud{In this work,} the magnetic origin of fan-like jets on the scale of inter-granule is studied in details for the first time. 
\ud{Magnetic cancellation near the inter-granular lanes is suggested to be responsible for the recurrent jets. Horizontal converging flows are identified, driving the repeatedly occurred cancellation. Doppler velocity of red shift is found to exist at the location of inter-granular lanes, indicating the reconnection downflow at the photosphere.}

\ud{The inter-granular lanes are located inside the region in between a group of sunspots, on one side of which there is a well developed and
isolated sunspot with umbra and penumbra. It is different from light bridge which commonly appears inside the umbra of a sunspot separating the umbra into two spots. Although the fan-like jets are not associated to a LB, the regions in between the sunspots show similar features, i.e. weak magnetic field
with same polarity compared to the ambient magnetic field.}
Comparing to the LB originated fan-like jets in \citet[][S09]{2009ApJ...696L..66S}, the jets studied in our case have a larger length in scale (10 Mm vs 1.3 Mm in S09), a \ud{similar speed (42 $\rm km\ s^{-1}$ vs 40 $\rm km\ s^{-1}$ in S09)}. The photospheric footpoints also show high value of Doppler velocity (2 $\rm km\ s^{-1}$ vs 0.73 $\rm km\ s^{-1}$ in S09) at the places where magnetic cancellation happens. No large-scale vertical current sheet like shown in S09 is found in our case but magnetic flux cancellation is identified. Therefore, a convection-driven magnetic cancellation on granular scale rather than a large-scale flux rope emergence is \ud{more plausible} to be the main cause of the magnetic reconnection at the footpoint that is responsible for the recurrent fan-like jets in this study. \\

The convective nature at the photosphere in our case is consistent with the simulation results of LB in \citet[][T15]{2015ApJ...811..138T} \ud{where} the convective upflow transports horizontal fields to the surface layers and creates magnetic configuration favorable for magnetic reconnection. The inter-granular lanes that generate fan-like jets have \ud{convective} flows with mixed magnetic polarities. The flows play a role to squash the opposite polarities and make the reconnection \ud{happens}. The inter-granular lanes are occupied with strong magnetic field on the magnitude of active region either along LOS or at the horizontal plane. This finding is complementary to the \ud{picture of granule kinematics in quiet region}, which assumes a $\rm  B_{l}$ dominated magnetic field at the inter-granule lanes. 
The Doppler velocity of the convective flow might be covered by the reconnection ouflow while the horizontal velocity ($\rm  V_x$ in T15) in our case is \ud{as fast as} 4 $\rm km\ s^{-1}$ -- larger than the observation results \citep{2015ApJ...811..137T} from Hinode but consistent with the simulation one. \\

Acknowledgements. \ud{We thank the anonymous referee for the valuable suggestions to greatly improve our paper.} We thank fruitful discussion with Prof. Pengfei Chen, Prof. Hui Tian, Dr. Zhi Xu and Dr. Ying Li.
\ud{J.Z. was a visiting postdoc at HAO, supported by the Chinese Scholarship Council (CSC NO. 201704910457). This work is also supported by Chinese Academy of Science Strategic Pioneer Program on Space Science, Grant No. XDA15052200, XDA15320103, XDA15320301 and by National Natural Science Foundation of China, Grant No. U1731241 and 11503089. 
BBSO operation is supported by NJIT and US NSF AGS-1821294 grant. GST operation is partly supported by the Korea Astronomy and Space Science Institute and the Seoul National University. X.Y. and K.A. acknowledge support from US NSF AST-2108235 and NASA 80NSSC20K0025 grants.}

\bibliographystyle{aasjournal} 
\bibliography{jets}

\section*{Figure captions}

\end{CJK*}
\end{document}